\documentclass[aps,pra,twocolumn,floatfix,nofootinbib]{revtex4}
\usepackage{amssymb}
\usepackage{bm}
\usepackage{epsfig}

\def\be{\begin{equation}}
\def\ee{\end{equation}}
\def\bea{\begin{eqnarray}}
\def\eea{\end{eqnarray}}

\begin{document}

\title{Cooperative Spontaneous Emission as a Many Body Eigenvalue Problem}
\author{Anatoly Svidzinsky and Jun-Tao Chang}
\affiliation{{\small Institute for Quantum Studies and Department of Physics, Texas A\&M
Univ., Texas 77843 }}
\date{\today }

\begin{abstract}
We study emission of a single photon from a spherically symmetric cloud of N
atoms (one atom is excited, N-1 are in ground state) and present an exact
analytical expression for eigenvalues and eigenstates of this many body
problem. We found that some states decay much faster then the single-atom
decay rate, while other states are trapped and undergo very slow decay. When
size of the atomic cloud is small compared with the radiation wave length we
found that the radiation frequency undergoes a large shift.
\end{abstract}

\maketitle

Recent quantum optical experiments and calculations \cite{Scul06a,Scul06b}
focus on the problem in which a single photon is stored in a gas cloud and
then retrieved at a later time. The directionality and spectral content of
the cooperatively reemitted photon is then of interest.

Furthermore synchrotron radiation experiments involving $N$ nuclei excited
by weak $\gamma $ ray pulse have features in common with the present problem 
\cite{Lipkin}. For example, in such experiments a thin disk of nuclei can
easily be prepared in a superposition in which the atoms are all in the
ground state together with a small probability of a uniform excitation of
the state, similar to Eq. (\ref{a0}), added in. The simplest example of
two-atom cooperative decay has been studied in many publications \cite%
{Step64}. The $N$-atom problem has been also investigated by several authors 
\cite{Bule64}. Time evolution and directionality of the radiation emitted
from a system of two-level atoms which are excited by a plane-wave pulse
have been discussed in \cite{Rehl71}.

Having motivated our interest in the problem we now turn to the analysis of
the correlated spontaneous emission from $N$ atoms in free-space. We
consider a system of two level ($a$ and $b$) atoms, initially one of them is
in the excited state $a$ and $E_{a}-E_{b}=\hslash \omega $. Initially there
are no photons. Atoms are located at positions $\mathbf{r}_{j}$ ($j=1,...,N$%
). In the dipole approximation the interaction of atoms with photons is
described by the Hamiltonian%
\begin{equation}
\hat{H}_{\text{int}}=\sum_{\mathbf{k}}\sum_{j=1}^{N}g_{k}\left[ \hat{\sigma}%
_{j}\hat{a}_{k}^{\dag }\exp (i(\nu _{k}-\omega )t-i\mathbf{k\cdot r}_{j})+%
\text{adj}\right] ,  \label{d1}
\end{equation}%
where $\hat{\sigma}_{j}$ is the lowering operator for atom $j$, $\hat{a}_{k}$
is the photon operator and $g_{k}$ is the atom-photon coupling constant for
the $k$ mode. We look for a solution of the Schr\"{o}dinger equation for the
atoms and the field as a superposition of Fock states%
\[
\Psi =\sum_{j=1}^{N}\beta _{j}(t)|b_{1}b_{2}...a_{j}...b_{N}>|\,0>+ 
\]%
\begin{equation}
\sum_{\mathbf{k}}\gamma _{\mathbf{k}}(t)|b_{1}b_{2}...b_{N}>|1_{\mathbf{k}}>.
\label{d2}
\end{equation}%
States in the first sum correspond to zero number of photons, while in the
second sum the photon occupation number is equal to one and all atoms are in
the ground state $b$. For simplicity we neglect the effects of photon
polarization. Substitute of Eq. (\ref{d2}) into the Schr\"{o}dinger equation
yields the following equations for $\beta _{j}(t)$ and $\gamma _{k}(t)$ (we
put $\hbar =1$) 
\begin{equation}
\dot{\beta}_{j}(t)=-i\sum_{\mathbf{k}}g_{k}\gamma _{k}(t)\exp [-i(\nu
_{k}-\omega )t+i\mathbf{k\cdot r}_{j}],  \label{d3}
\end{equation}%
\begin{equation}
\dot{\gamma}_{\mathbf{k}}(t)=-i\sum_{j=1}^{N}g_{k}\beta _{j}(t)\exp [i(\nu
_{k}-\omega )t-i\mathbf{k\cdot r}_{j}].  \label{d4}
\end{equation}%
Integrating Eq. (\ref{d4}) over time gives [$\gamma _{k}(0)=0$]%
\begin{equation}
\gamma _{\mathbf{k}}(t)=-i\int_{0}^{t}dt^{\prime }\sum_{j=1}^{N}g_{k}\beta
_{j}(t^{\prime })\exp [i(\nu _{k}-\omega )t^{\prime }-i\mathbf{k\cdot r}%
_{j}].  \label{d5}
\end{equation}%
Substituting this into (\ref{d3}) we obtain equation for $\beta _{j}(t)$ 
\begin{equation}
\dot{\beta}_{j}(t)=-\sum_{\mathbf{k}}\sum_{j^{\prime
}=1}^{N}\int_{0}^{t}dt^{\prime }g_{k}^{2}\beta _{j^{\prime }}(t^{\prime
})e^{i(\nu _{k}-\omega )(t^{\prime }-t)+i\mathbf{k\cdot (r}_{j}-\mathbf{r}%
_{j^{\prime }})}.  \label{i2}
\end{equation}%
We proceed by making the Markov approximation a-la Weisskopf and Wigner to
obtain%
\begin{equation}
\dot{\beta}_{i}(t)=-\gamma \sum_{j=1}^{N}\Gamma _{ij}\beta _{j}(t),
\label{q2}
\end{equation}%
where for $i\neq j$ 
\begin{equation}
\Gamma _{ij}=\frac{\sin (k_{0}|\mathbf{r}_{i}-\mathbf{r}_{j}|)}{k_{0}|%
\mathbf{r}_{i}-\mathbf{r}_{j}|}-i\frac{\cos (k_{0}|\mathbf{r}_{i}-\mathbf{r}%
_{j}|)}{k_{0}|\mathbf{r}_{i}-\mathbf{r}_{j}|},  \label{q3}
\end{equation}%
$\Gamma _{ii}=1$, $k_{0}=\omega /c$ and $\gamma $ is the single atom
spontaneous decay rate

\[
\gamma =\frac{V_{\text{ph}}k_{0}^{2}g_{k_{0}}^{2}}{\pi c}, 
\]%
$V_{\text{ph }}$is the photon volume.

We point out that a rigorous treatment of the problem beyond the rotating
wave approximation Hamiltonian (\ref{d1}) also yields Eqs. (\ref{q2}) and (%
\ref{q3}) \cite{Lehm70,Mana73}. Imaginary part of $\Gamma _{ij}$ in Eq. (\ref%
{q3}) appears due to a short range interaction between atoms which is
induced by electromagnetic field and causes a frequency shift \cite{Frie74}.
The frequency shift becomes substantial when size of the atomic cloud is
smaller then the wave length, this will be clear from Eq. (\ref{c38}) below.

One can rewrite Eq. (\ref{q2}) in a matrix form%
\begin{equation}
\dot{B}=-\gamma \Gamma B,  \label{d10}
\end{equation}%
where the vector $B$ and the decay matrix $\Gamma $ are given by%
\begin{equation}
B=\left( 
\begin{array}{c}
\beta _{1}(t) \\ 
\beta _{2}(t) \\ 
\vdots \\ 
\beta _{N}(t)%
\end{array}%
\right) ,\quad \Gamma =||\Gamma _{ij}||=\left( 
\begin{array}{cccc}
1 & \Gamma _{12} & \cdots & \Gamma _{1N} \\ 
\Gamma _{21} & 1 & \cdots & \Gamma _{2N} \\ 
\vdots & \vdots & \ddots & \vdots \\ 
\Gamma _{N1} & \Gamma _{N2} & \cdots & 1%
\end{array}%
\right) .  \label{d11}
\end{equation}
The matrix $\Gamma $ is symmetric $\Gamma _{ij}=\Gamma _{ji}$. Let $|\lambda
_{i}>$ be eigenvectors of $\Gamma $ and $\lambda _{i}$ ($i=1,...,N$) are the
corresponding eigenvalues. If initially the system is prepared in an
eigenstate $B(0)=|\lambda _{i}>$, then according to Eq. (\ref{d10}) the
state evolution is given by 
\begin{equation}
B(t)=e^{-\gamma \lambda _{i}t}B(0),
\end{equation}%
that is state evolves independently of other states and decays with the rate 
$\gamma $Re$\lambda _{i}$, where $\gamma $ is the spontaneous decay rate of
a single atom. The general solution of the Schr\"{o}dinger equation is%
\[
\Psi =C_{1}e^{-\gamma \lambda _{1}t}|\lambda _{1}>+C_{2}e^{-\gamma \lambda
_{2}t}|\lambda _{2}>+\ldots + 
\]%
\begin{equation}
C_{N}e^{-\gamma \lambda _{N}t}|\lambda _{N}>+\sum_{\mathbf{k}}\gamma _{%
\mathbf{k}}(t)|b_{1}b_{2}...b_{N}>|1_{\mathbf{k}}>,
\end{equation}%
where $C_{1}$, $C_{2}$, $...$, $C_{N}$ are constants determined by the
initial conditions. Real part of $\lambda _{i}$ are positive numbers and,
hence, the general solution corresponds to an exponential decay of the
initial state.

One should note that since tr$(\Gamma )=N$ we obtain%
\begin{equation}
\sum_{i=1}^{N}\lambda _{i}=\text{tr}(\Gamma )=N,
\end{equation}%
which is useful and insightful result.

Next we solve the eigenvalue Eq. (\ref{q2}) analytically for a dense
spherically symmetric cloud with atomic density $\rho (r)$ ($\int \rho (r)d%
\mathbf{r}=N$). $\Gamma _{ij}$ changes in a scale of $1/k_{0}$. Assuming
there are many atoms in the volume $1/k_{0}^{3}$ we can replace summation by
integration, then Eq. (\ref{q2}) reads ($\beta =e^{-\gamma \lambda t}\beta (%
\mathbf{r})$)%
\begin{equation}
\int d\mathbf{r}^{\prime }\rho (r^{\prime })\left[ K_{0}(\mathbf{r},\mathbf{r%
}^{\prime })+iK_{1}(\mathbf{r},\mathbf{r}^{\prime })\right] \beta (\mathbf{r}%
^{\prime })=\lambda \beta (\mathbf{r}),  \label{f1}
\end{equation}%
where%
\[
K_{0}(\mathbf{r},\mathbf{r}^{\prime })=\frac{\sin (k_{0}|\mathbf{r}-\mathbf{r%
}^{\prime }|)}{k_{0}|\mathbf{r}-\mathbf{r}^{\prime }|},\quad K_{1}(\mathbf{r}%
,\mathbf{r}^{\prime })=-\frac{\cos (k_{0}|\mathbf{r}-\mathbf{r}^{\prime }|)]%
}{k_{0}|\mathbf{r}-\mathbf{r}^{\prime }|}. 
\]

\begin{figure}[h]
\bigskip 
\centerline{\epsfxsize=0.55\textwidth\epsfysize=0.45\textwidth
\epsfbox{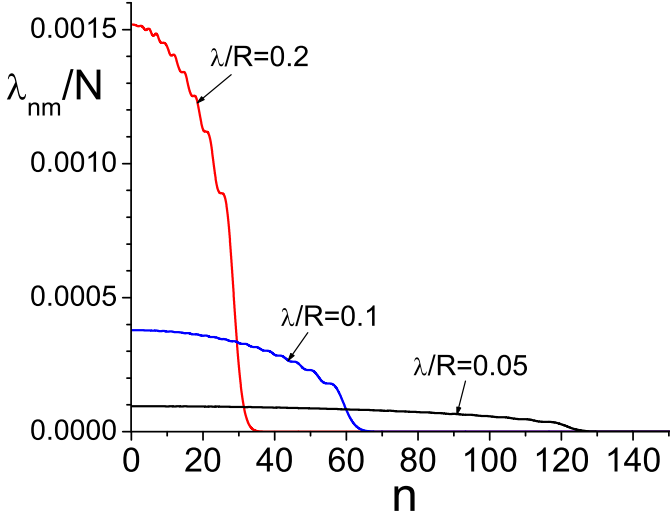}}
\caption{$\protect\lambda _{nm}/N$ as a function of $n$ obtained from Eq. (%
\protect\ref{f24}) at different $\protect\lambda /R$. }
\label{la3}
\end{figure}

It turns out that for $k_{0}R\gg 1$, where $R$ is the characteristic size of
atomic cloud, the eigenfunctions of the integral Eq. (\ref{f1}) are
determined by the real part of the kernel, $K_{0}(\mathbf{r},\mathbf{r}%
^{\prime })$, that is by equation%
\begin{equation}
\int d\mathbf{r}^{\prime }\rho (r^{\prime })K_{0}(\mathbf{r},\mathbf{r}%
^{\prime })\beta (\mathbf{r}^{\prime })=\lambda \beta (\mathbf{r}).
\label{f1a}
\end{equation}%
Eq. (\ref{f1a}) can be solved using the identity \cite{Grad80}%
\[
\frac{\sin (k_{0}|\mathbf{r}-\mathbf{r}^{\prime }|)}{k_{0}|\mathbf{r}-%
\mathbf{r}^{\prime }|}=4\pi \sum_{n=0}^{\infty
}\sum_{m=-n}^{n}j_{n}(k_{0}r)Y_{nm}(\hat{r})Y_{nm}^{\ast }(\hat{r}^{\prime
})j_{n}(k_{0}r^{\prime }) 
\]%
and the orthogonality condition%
\[
\int d\Omega _{r}Y_{nm}^{\ast }(\hat{r})Y_{ks}(\hat{r})=\delta _{nk}\delta
_{ms}, 
\]%
where $Y_{nm}$ are spherical harmonics, $\hat{r}$ is a unit vector in the
direction of $\mathbf{r}$ and $j_{n}(x)$ are the spherical Bessel functions.
Solutions of Eq. (\ref{f1a}) are%
\begin{equation}
\beta _{nm}(\mathbf{r})=j_{n}(k_{0}r)Y_{nm}(\theta ,\varphi ),  \label{f2}
\end{equation}

\begin{equation}
\lambda _{nm}=\int d\mathbf{r}\rho (r)j_{n}^{2}(k_{0}r),  \label{f25}
\end{equation}%
where $\theta $ and $\varphi $ are angles describing direction of $\mathbf{r}
$ in the spherical coordinate system. Eigenvalues (\ref{f25}) are
independent of $m$. That is the eigenvalues $\lambda _{nm}$ are $(2n+1)$%
-fold degenerate. If atoms are uniformly distributed inside a sphere of
radius $R$ (that is $\rho (r)=N/V$, for $r<R$) we obtain \cite{Erns69}%
\begin{equation}
\lambda _{nm}=\frac{3N}{2}\left[
j_{n}^{2}(k_{0}R)-j_{n-1}(k_{0}R)j_{n+1}(k_{0}R)\right] .  \label{f24}
\end{equation}%
In the limit $k_{0}R\gg n$ we find

\begin{equation}
\beta _{nm}(\mathbf{r})\approx \frac{1}{r}\sin \left( k_{0}r-\frac{\pi }{2}%
n\right) Y_{nm}(\theta ,\varphi ),
\end{equation}%
\begin{equation}
\lambda _{nm}\approx \frac{3N}{2(k_{0}R)^{2}}.  \label{s23}
\end{equation}%
Fig. \ref{la3} shows $\lambda _{nm}/N$ as a function of $n$ obtained from
Eq. (\ref{f24}) at different $\lambda /R\ll 1$ ($\lambda =2\pi /k_{0}$). The
states with $n<k_{0}R$ are degenerate and decay with the rate $3\gamma
N/2(k_{0}R)^{2}$, while decay of the states with $n>k_{0}R$ is suppressed.

$K_{1}(\mathbf{r},\mathbf{r}^{\prime })$ term in Eq. (\ref{f1}) yields an
imaginary contribution in $\lambda _{nm}$ (frequency shift) and also
modifies the real part. To solve Eq. (\ref{f1}) we use the identity \cite%
{Grad80}%
\[
\frac{\exp (ik_{0}|\mathbf{r}-\mathbf{r}^{\prime }|)}{k_{0}|\mathbf{r}-%
\mathbf{r}^{\prime }|}=4\pi i\sum\limits_{k=0}^{\infty
}\sum_{s=-k}^{k}Y_{ks}(\hat{r})Y_{ks}^{\ast }(\hat{r}^{\prime }) 
\]%
\begin{equation}
\times \left\{ 
\begin{array}{c}
j_{k}(k_{0}r^{\prime })h_{k}^{(1)}(k_{0}r),\quad r>r^{\prime } \\ 
j_{k}(k_{0}r)h_{k}^{(1)}(k_{0}r^{\prime }),\quad r\leq r^{\prime }%
\end{array}%
\right. ,  \label{c2}
\end{equation}%
where $h_{k}^{(1)}(z)$ are the spherical Bessel functions, $h_{k}^{(1)}(z)=%
\sqrt{\pi /2z}H_{k+1/2}^{(1)}\left( z\right) $, $H_{k+1/2}^{(1)}\left(
z\right) $ are the Hankel functions of the first kind.

If atoms are uniformly distributed inside a sphere of radius $R$ the answer
for eigenfunctions is given by%
\begin{equation}
\beta (\mathbf{r})=j_{n}\left( ak_{0}r\right) Y_{nm}(\hat{r}),  \label{c27a}
\end{equation}%
where%
\begin{equation}
a=\sqrt{1-\frac{3Ni}{k_{0}^{3}R^{3}\lambda _{n}}},  \label{c22}
\end{equation}%
and the eigenvalues $\lambda _{n}$ are determined from the following
equation for $a$%
\begin{equation}
a=\frac{j_{n}\left( ak_{0}R\right) }{j_{n-1}\left( ak_{0}R\right) }\frac{%
h_{n-1}^{(1)}(k_{0}R)}{h_{n}^{(1)}(k_{0}R)}.  \label{c27}
\end{equation}%
In the Dicke limit $k_{0}R\ll 1$ Eqs. (\ref{c27a}) and (\ref{c27}) yield 
\begin{equation}
\beta _{nlm}(\mathbf{r})=j_{n}\left( A_{nl}\frac{r}{R}\right) Y_{nm}(\hat{r}%
),  \label{f2a}
\end{equation}%
\begin{equation}
\lambda _{nl}\approx -\frac{3iN}{A_{nl}^{2}k_{0}R}+\frac{6N(k_{0}R)^{2n}}{%
A_{nl}^{4}[(2n-1)!!]^{2}},  \label{c38}
\end{equation}%
where $A_{nl}$ are nonnegative zeroes of the Bessel function $j_{n-1}\left(
x\right) $. In particular, $A_{0l}=(2l-1)\pi /2$ and $A_{1l}=\pi l$, $%
l=1,2,3\ldots $. One can see that $\text{Im}(\lambda _{nl})$ (frequency
shift) becomes large for $k_{0}R\rightarrow 0$. In the Dicke limit only
eigenvalues with $n=0$ have large real part and decay fast with the rate $%
\Gamma _{l}=96N\gamma /\pi ^{4}(2l-1)^{4}$ (Dicke superradiance \cite{Dick54}%
), while eigenvalues with $n>0$ are suppressed by a factor $(k_{0}R)^{2n}$.
Those states are trapped. Please note that $\sum_{l=1}^{\infty }\Gamma
_{l}=N\gamma $, as expected from general arguments. To check validity of our
analytical results we solved numerically the eigenvalue problem for the
matrix (\ref{d11}) with atoms randomly distributed inside a sphere. In the
Dicke limit our numerical simulations show excellent agreement with Eq. (\ref%
{c38}) already for a few hundred atoms.

In the limit $k_{0}R\gg n$ we find%
\[
\lambda _{n}\approx \frac{3N}{2(k_{0}R)^{2}}\left[ 1-\frac{(-1)^{n}}{2k_{0}R}%
\sin (2k_{0}R)\right. 
\]%
\begin{equation}
-\left. i\frac{(-1)^{n}}{2k_{0}R}(\cos (2k_{0}R)-1)\right] .  \label{c25}
\end{equation}%
In such a limit the contribution from the $K_{1}$ term in the kernel is
smaller by a factor $1/k_{0}R$ then those from the $K_{0}$ piece.

\begin{figure}[h]
\bigskip 
\centerline{\epsfxsize=0.5\textwidth\epsfysize=0.4\textwidth
\epsfbox{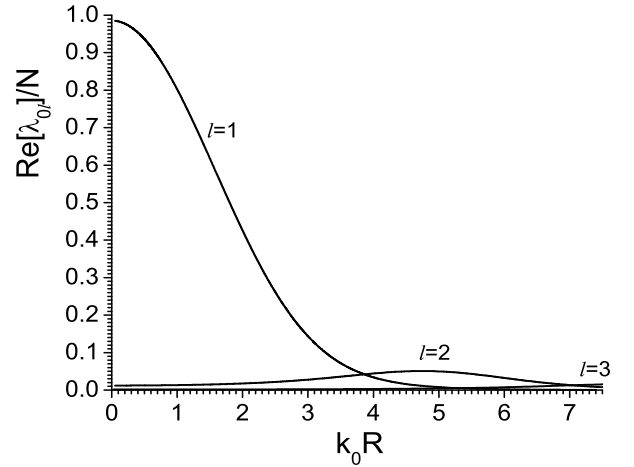}}
\caption{Real part of $\protect\lambda_{0l}$ as a function of $k_0R$ for $%
l=1 $, $2$ and $3$.}
\label{rel0}
\end{figure}

\begin{figure}[h]
\bigskip 
\centerline{\epsfxsize=0.5\textwidth\epsfysize=0.4\textwidth
\epsfbox{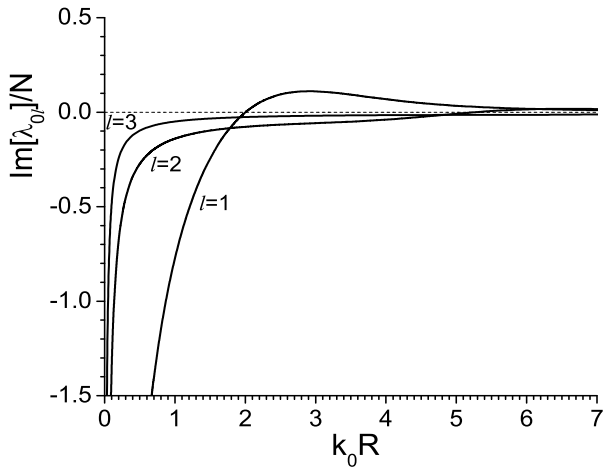}}
\caption{Imaginary part of $\protect\lambda_{0l}$ as a function of $k_0R$
for $l=1$, $2$ and $3$.}
\label{iml0}
\end{figure}

Figs. \ref{rel0} and \ref{iml0} show real and imaginary part of $\lambda
_{0l}$ as a function of $k_{0}R$ obtained by solving Eq. (\ref{c27})
numerically.

We next discuss the angular distribution of the emitted radiation produced
by an eigenstate $\beta _{nlm}.$ Taking into account $\beta
_{j}(t)=e^{-\gamma \lambda _{nl}t}\beta _{nlm}(\mathbf{r}_{j})$ we obtain
from Eq. (\ref{d5}) for a uniform cloud of radius $R$%
\[
\gamma _{\mathbf{k}}(t)=g_{\mathbf{k}}\frac{N}{V}\frac{\left[ 1-e^{-\gamma
\lambda _{nl}t+i(\nu _{k}-\omega )t}\right] }{\omega -\nu _{k}+i\gamma
\lambda _{nl}}\times 
\]%
\begin{equation}
\int_{0}^{R}drr^{2}\int d\Omega _{r}\beta _{nlm}(\mathbf{r})\exp [-i\mathbf{%
k\cdot r}].
\end{equation}%
Substitute Eq. (\ref{c27a}) yields $\gamma _{\mathbf{k}}(t)\propto Y_{nm}(%
\hat{k})$. That is the angular distribution of the emitted radiation is
given by the same spherical function $Y_{nm}(\theta ,\varphi )$ which
describes anisotropy of the initial eigenstate.

Let us next apply the proceeding to the problem of a uniformly excited
sample in which one atom is excited and the state is given by%
\begin{equation}
|\Psi (0)\rangle =\frac{1}{\sqrt{N}}\sum\limits_{j=1}^{N}e^{i\mathbf{k}%
_{0}\cdot \mathbf{r}_{j}}|b_{1},b_{2},...,a_{j},...,b_{N}\rangle |0\rangle .
\label{a0}
\end{equation}%
This is an interesting many body problem comprised of $N$ atoms collectively
emitting a single photon \cite{Scul06a}. For example, this state can be
prepared by the experiments of the type described in Ref. \cite{Scul06b}. We
are interested in the state of the system at time $t$ in the limit $%
k_{0}R\gg 1$. We obtain the state evolution by expanding $|\Psi (0)\rangle $
in terms of the eigenvectors $|\lambda _{nm}>$%
\begin{equation}
|\lambda _{nm}>=\sum\limits_{j=1}^{N}j_{n}(k_{0}r_{j})Y_{nm}(\hat{r}%
_{j})|b_{1},b_{2},...,a_{j},...,b_{N}\rangle |0\rangle .
\end{equation}%
We do this using the relation

\begin{equation}
\exp (i\mathbf{k}_{0}\cdot \mathbf{r}_{j})=4\pi \sum_{n=0}^{\infty
}\sum_{m=-n}^{n}i^{n}j_{n}(k_{0}r_{j})Y_{nm}^{\ast }(\hat{k}_{0})Y_{nm}(\hat{%
r}_{j}),  \label{f8}
\end{equation}%
where $\hat{k}_{0}$ and $\hat{r}_{j}$ are unit vectors in the directions $%
\mathbf{k}_{0}$ and $\mathbf{r}_{j}$ respectively. The state (\ref{a0})
evolves as%
\begin{equation}
|\Psi (t)\rangle _{\text{atomic}}=\frac{4\pi }{\sqrt{N}}\sum_{n=0}^{\infty
}\sum_{m=-n}^{n}i^{n}Y_{nm}^{\ast }(\hat{k}_{0})e^{-\lambda _{n}\gamma
t}|\lambda _{nm}>
\end{equation}%
where $\lambda _{n}\gamma \approx 3\gamma N/2(k_{0}R)^{2}\equiv \Gamma _{N}$%
, see Eq. (\ref{c25}). Hence we have for the atomic state%
\begin{equation}
|\Psi (t)\rangle _{\text{atomic}}=e^{-\Gamma _{N}t}|\Psi (0)\rangle
\end{equation}%
and therefore%
\begin{equation}
\beta _{j}(t)=\frac{1}{\sqrt{N}}e^{-\Gamma _{N}t}e^{i\mathbf{k}_{0}\mathbf{%
\cdot r}_{j}}.
\end{equation}%
Substituting this into (\ref{d5}) we obtain%
\begin{equation}
\gamma _{\mathbf{k}}(t)=\frac{g_{k}}{\sqrt{N}}\frac{\left[ \exp \left(
-\Gamma _{N}t+i(\nu _{k}-\omega )t\right) -1\right] }{\omega -\nu
_{k}-i\Gamma _{N}}\sum_{j=1}^{N}e^{-i(\mathbf{k}_{0}-\mathbf{k)\cdot r}_{j}},
\label{f33}
\end{equation}%
which yields the following probability of a photon being emitted with the
wave vector $\mathbf{k}$ \cite{Scul06b,Rehl71}%
\begin{equation}
|\gamma _{\mathbf{k}}(\infty )|^{2}=\frac{g_{k}^{2}}{N}\frac{1}{(\omega -\nu
_{k})^{2}+\Gamma _{N}^{2}}\left\vert \sum_{j=1}^{N}e^{-i(\mathbf{k}_{0}-%
\mathbf{k)\cdot r}_{j}}\right\vert ^{2}.
\end{equation}

The decay rate $\Gamma _{N}$ can be understood as the following. Dicke-like
arguments \cite{Dick54} for a coherent decay would yield the decay rate of $%
N\gamma $. However, the spontaneous decay rate is due to emission in all $%
4\pi $ sr directions, while the state (\ref{a0}) emits photon in a small
diffraction angle $\lambda /R$. As a result, the Dicke superradiance rate $%
N\gamma $ is reduced by the ratio of the solid diffraction angle $\lambda
^{2}/R^{2}$ to $4\pi $ sr, this yields $\Gamma _{N}$.

In summary, we studied correlated spontaneous emission from $N$ atoms in a
spherically symmetric cloud which is a $N-$body problem. For a dense atomic
gas we obtained analytical expression for the eigenstates and eigenvalues of
the system. We found that some states decay much faster then the single-atom
decay rate, while other states are trapped and undergo very slow decay. In
the limit $\lambda \gg R$ only states with $n=0$ decay fast ($\Gamma
_{N}\sim N\gamma $), all other states are trapped. In the opposite limit $%
\lambda \ll R$ many states decay with the same rate $\Gamma _{N}\approx
N\gamma \lambda ^{2}/R^{2}$. We also found that for $\lambda \gg R$ the
eigenvalues have a large imaginary part which corresponds to a frequency
shift of emitted radiation.

We are very grateful to R. Glauber, P. Berman, R. Friedberg, J.T. Manassah,
S. Prasad and M.O. Scully for valuable discussions. This work was supported
by the Office of Naval Research (Award No. N00014-03-1-0385) and the Robert
A. Welch Foundation (Grant No. A-1261).


\begin{thebibliography}{99}
\bibitem{Erns69} V. Ernst, Z. Physik \textbf{218}, 111 (1969). It is
important to emphasize that Ernst's paper considers Eq. (\ref{f1a}) which
has incomplete kernel and is valid only for $k_{0}R\gg 1$. Here we study Eq.
(\ref{f1}) which is valid for any $k_{0}R$.
\end{thebibliography}
\end{document}